# Topological Insulators in Hexagonal Wurtzite-type Binary Compounds

X. M. Zhang, [1] R S. Ma, [1] X. C. Liu, [1] E. K. Liu, [1] G. D. Liu, [2] Z. Y. Liu,[3] W. H. Wang, [1, a)] and G. H. Wu[1]

[1]*Beijing National Laboratory for Condensed Matter Physics, Institute of Physics, Chinese Academy of Sciences, Beijing 100080, P. R. China*

[2] *School of Material Sciences and Engineering, Hebei University Technology, Tianjin 300130, P. R. China*

[3]*State Key Laboratory of Metastable Material Sciences and Technology, Yanshan University Technology, Qinhuangdao 066004, P. R. China*

**Abstract:** We propose new topological insulators in hexagonal wurtzite-type binary compounds based on the first principles calculations. It is found that two compounds AgI and AuI are three-dimensional topological insulators with a naturally opened band-gap at Fermi level. From band inversion mechanism point view, this new family of topological insulators is similar with HgTe, which has $s$ ($\Gamma_6$) − $p$ ($\Gamma_8$) band inversion. Our results strongly support that the spin-orbit coupling is not an essential factor to the band inversion mechanism; on the contrary, it is mainly responsible to the formation of a global band gap for the studied topological insulators. We further theoretically explore the feasibility of tuning the topological order of the studied compounds with two types of strains. The results show that the uniaxial strain can contribute extremely drastic impacts to the band inversion behavior, which provide an effective approach to induce topological phase transition.

## I. INTRODUCTION

Topological insulators (TIs) are newly discovered materials with an insulating bulk band gap and topologically protected metallic surface states. [1-4] The first theoretical predications on binary Bi-based TIs, such as $Bi_2Se_3$ and $Bi_2Te_3$, have already been well studied and experimentally realized. [5-11] Recently, the search for TIs has been extended to ternary Heusler and chalcopyrite compounds, [12-16] and also the layered honeycomb lattice $Na_2IrO_3$. [17] To obtain a topological nontrivial phase, band inversion plays an indispensable role for turning topological order. Based on the band inversion mechanism, TIs can be classified into two types: [18] (1) HgTe-like TIs, where the band inversion occurs between the $S_+$ and $P_-$ states. (2) $Bi_2Se_3$-like TIs, in which $P_{z,+}$ and $P_{z,-}$ orbitals are inverted. Generally, TIs with cubic or tetragonal structure are more likely to follow the former category, while the hexangular one tends to obey the latter. Previously, the spin-orbit coupling (SOC) seems to describe very well the reason of the band inversion. [1,4] While the recent works point out that the scalar relativistic (SR) effects and/or lattice distortion is the root cause, [18,19] which provide a deeper understanding to the origin of band inversion.

Although a large number of TIs have been predicted and thoroughly studied, the efforts for exploiting new TIs is consistently undiminished. The cubic zinc-blende compound HgTe, which is a well studied two-dimensional (2D) TI, [20,21] has given people much inspiration to trace new TIs. Typically, the XYZ half-Heuslers can be viewed as an $X^{n+}$ ion 'stuffing' the zinc-blende $YZ^{n-}$ sublattice to illustrate their topological nontrivial behavior. [12] Analogous thoughts also give the birth to the TIs in chalcopyrite, honeycomb and inverse-Heusler families. [16,22,23] Similar to the topological nontrivial phase HgTe, many III-V, II-VI and I-VII semiconductors crystalize the cubic zinc-blende structure, as shown in Fig. 1 (a) and (c). However, some of them, such as GaAs, ZnS and AgI can also form a hexangular wurtzite phase [see Fig. 1(b) and (d)] under proper pressure or temperature. [24-26] Here, we select the hexangular wurtzite compounds for exploring TIs based on the following facts: (i) Wurtzite and zinc-blende compounds are expected to possess some common natural instincts, since they are the allotropes for most of the above given semiconductors; (ii) there exist some overall similarity between the zinc-blende and wurtzite structure: both of them yield a four-coordinated lattice, in which each atom is environed by four heterogeneous ones as nearest neighbours. Moreover, (iii) the cubic zinc-blende supercell can exhibit the sight of hexagonal lattice along the (111) direction [see Fig. 1 (c)], which is comparable to the top side view of wurtzite in Fig. 1 (d).

In this paper, we perform a systematic investigation on the band topology of the wurtzite compounds AgI and AuI. With accredited deduction and comparison to the well studied zinc-blende HgTe, we show that the hexagonal wurtzite AuI and slightly strained-AgI are naturally band-gapped HgTe-like topological insulators, which is distinct from the previous hexagonal $Bi_2Si_3$-like TIs. Our finding further support that SR effects and strain are prominent to the downward shift of bands with $\Gamma_6$ symmetry, while the SOC gives the rise of a global band gap. Finally, we give an exploration to the strain effects on turning the topological order.

## II. METHODOLOGY

The band-structure calculations are performed using full-potential linearized augmented plane-wave method implemented in the WIEN2K package, [27,28] which outputs highly accurate results for inhomogeneous structures. A converged ground state was obtained using 5 000 k points in the first Brillouin zone with $K_{max} * R_{MT} = 9.0$, where $R_{MT}$ represents the muffin-tin radius and $K_{max}$ is the maximum size of the reciprocal-lattice vectors. Wave functions and potentials inside the atomic sphere are expanded in spherical harmonics up to l=10 and 4, respectively. The spin-orbit coupling(SOC) is treated by means of the second variational procedure with the scalar-relativistic orbitals as basis, where states up to 10 Ry above the Fermi level are included in the basis expansion.

We use experimental lattice constant for the unstrained structure of Wurtzite-type AgI (a=b=4.59Å, c=7.51Å) and an equilibrium one determined by total energy minimization for AuI ($a_{eq}$= 4.64 Å, $c_{eq}$ =7.56 Å). [26] It should be pointed out that, even though AuI has not been experimentally synthesized in wurtzite structure, it would primely serve our theoretical study as a counterpart to the AgI. Simulated by variation of lattice, we perform two types of strain to test the stability of the topological insulating behavior. One is the hydrostatic strain, which can be achieved by equally increasing or decreasing the lattice constants along all the three axes. Such condition provides more of academic significance rather than for practical applications. The other type of strain is uniaxial one, realized by varying the lattice constant in the ab plane within a constant cell volume. This gives good simulation of the growth for thin films.

## III. RESULTS AND DISCUSSION

Figure 2 shows the detail band structures of the zinc-blende (a) CdTe and (d) HgTe and make a comparation with those of wurtzite compounds (b, c) AgI and (e, f) AuI. As shown in Fig. 2(a) and (d), the calculated band structures of the zinc-blende CdTe and HgTe are consistent with previous results,[29] which indicate CdTe and HgTe are a topological trivial and a nontrivial phase, respectively. As to the wurtzite compounds, we find the band structure of AgI has a natural band ordering [$\Gamma_6$ (red lines) $\rightarrow \Gamma_8$ (blue lines) $\rightarrow \Gamma_7$ (green lines) from top to bottom] and opens a direct gap around the Fermi level [see Fig.2(b)], which indicates it is a CdTe-like trivial phase; whereas the AuI exhibits a distinctly inverted band order [see Fig. 2(e)], where the $\Gamma_6$ states are occupied and sit below the $\Gamma_8$ ones. Analogous to the HgTe case, such a band inversion only occurs once at the $\Gamma$ point throughout the Brillouin zone and illuminates it a topological nontrivial candidate. However, unlike the HgTe-like TIs, as shown in the inset of Fig. 2 (e) for the wuritize AuI, the quartet degeneracy of $\Gamma_8$ states splits into two separated doublets and gives rise to a global energy gap of 63 meV, which is comparable to the well-known $Bi_2Se_3$ series [文献]. We should point out that, considering the naturally topological insulating behavior, the wurtzite AuI retains the merits of other TIs in hexangular system. [1,19,22] However, as we will discuss in the next paragraph that, the wuritze AuI exhibits the topological order switching between S and P bands, which is more of HgTe-like, rather than $Bi_2Se_3$-like from the band inversion point of view. Most interestingly, we found that the AuI displays the band inversion behavior even without considering SOC [see Fig. 2 (f)]. Combining with the comparation of the band structures for AgI with and without SOC [see Fig. 2(b) and Fig. 2(c)], we can conclude the effects of the SOC on reconstructing the bands near the $E_F$: (1) Split-off the degeneracy of corresponding bands. As shown in Fig. 2 (c) and (e), the individual $\Gamma_6$ (red line) and $\Gamma_7$ (green line) states splits into doublets after switching on the SOC. Similarly the two single (but overlapping at the $\Gamma$ point) bands with $\Gamma_8$ symmetry products a couple of separated doublets, which open up the global topological insulating gap at the Fermi level for AuI. (2) Narrow the energy difference $E_{BIS}=E_{\Gamma 6} - E_{\Gamma 8}$. The value of $E_{BIS}$ contributes as a crucial index for TIs which would be positive for topologically trivial cases and negative for the nontrivial ones. After turning on the SOC, the value of $E_{BIS}$ decreases from 1.387 eV to 1.142eV and -0.336eV to -0.509 eV for AgI and AuI, respectively. Our results further support the description that the topological phase transition can be induced by purely turning the SOC strength in ref. 12.

To better explain the mechanisms of the band inversion and the parity exchange for studied compounds in this work, in Fig. 3(a)-(d) we display the schematic diagrams of band evolution near the $\Gamma$ point. Starting from the atomic S and $P_{x,y,z}$ orbitals, we consider subsequently three procedures: switching on chemical bonding [procedure (I)], taking into account crystal field splitting [procedure (II)], and introducing SOC [procedure (III)]. In procedure (I), the chemical bonding hybridizations between the transition elements (Cd, Ag, Hg, Au) and p blocks (Te, I) lift up all of the S orbitals and push down the $p_{x,y,z}$ ones. The resulting coupled hybridizing states are given as $S_\pm$ and $P_{xyz,\pm}$, where "±" are the party labels. Subsequently, we take into account the effects of crystal field in procedure (II). For the zinc-blende CdTe and HgTe part [Fig 3.(a) and (c)], no further splittings are observed owning to the cubic symmetry, but accompanied with a decline of the corresponding energy levels, in particular for the S orbitals in HgTe, possessing a band inversion between $S_{Hg,+}$ and $P_{Te,xyz,-}$. On the other hand, for the wurtzite AgI and AuI cases [Fig 3.(b) and (d)], the $P_{I,xyz}$ orbitals split into $P_{I,z}$ and $P_{I,xy}$ owing to the hexagonal symmetry. But the non-splitting S orbitals experience similar energy declining with the CdTe and HgTe cases. And the downward of the S orbital is more severe for the heavier Au than Ag atom, leaving AuI with a inverted band order. So far, we can qualitatively explain the corresponding band structures for calculations without SOC shown in Fig 2(c) and (f) for AgI and AuI. The combining of procedure (I) and procedure (II) is known as SR effect.[30] Finally, in procedure (III), the SOC is taken into account. Since the aeolotropism introduced by the SOC, even the $P_{Te,xyz}$ orbital of the cubic CdTe and HgTe could be slightly splited. Moreover, subjecting to the atomic SOC, the $P_{xy}$, $P_z$ orbitals and the non-splitted S one will get to the derivative states $P^{\uparrow\downarrow}_{x+iy}$, $P^{\uparrow\downarrow}_{x-iy}$, $P^{\uparrow\downarrow}_z$, $S^{\uparrow\downarrow}$. We now obtain the bands with toplogical significance: the $P^\uparrow_{x+iy}$, $P^\downarrow_{x-iy}$ and $P^\downarrow_{x+iy}$. $P^\uparrow_{x-iy}$ corresponds to the two couples of bands with Γ8 symmetry, and $P^{\uparrow\downarrow}_z$, $S^{\uparrow\downarrow}$ are for those of $\Gamma_7$, $\Gamma_6$ symmetries. It should be point out that, for the system with the cubic symmetry, such as CdTe and HgTe, the slight disturbance caused by SOC is normally not enough to split off the $P^\uparrow_{x+iy}$, $P^\downarrow_{x-iy}$ and $P^\downarrow_{x+iy}$, $P^\uparrow_{x-iy}$ at the Γ point and induce HgTe a gapless TI. We show the $\Gamma_{8+}$ and $\Gamma_{8-}$ within separated energy levels in Fig. 3(a) and (c) just for clarity purpose. Differently, for the hexagonal AgI and AuI with lower crystal symmetry, the SOC breaks off the degeneracy between $\Gamma_{8+}$ and $\Gamma_{8-}$ at the $\Gamma$ point, yields a SOC-dependent gap in AuI. Similar effects of SOC are also found in ref. 19.

It has been well accepted that, strain plays an important role to the band topology for TIs. Especially, a recent search model for TIs has suggested that the variational 'strain descriptor' can be associated with the robustness or the feasibility of the TI state.[31] In the following, we will test the effects of the strain on tuning the topological order in the wurtzite AgI and AuI compounds. Two types of strain, namely hydrostatic and uniaxial, are simulated by varying the lattice constants and the datails have been shown in the 'METHODOLOGY' section. In Fig. 4 we show (a) the hydrostatic and (b) uniaxial strain-dependent $E_{BIS}$ as a function of the $a/a_0$ ratio for AgI. Analogous schemes for hydrostatic and uniaxial strain of AuI are shown in Fig. 4 (c) and (d). The calculated $E_{BIS}$ without and with SOC are griven as $E_{BIS}^{noSOC}$ and $E_{BIS}^{SOC}$, and their energy discrepancy is defined as $\Delta E_{BIS} = E_{BIS}^{noSOC} - E_{BIS}^{SOC}$, which indicates the degree of $E_{BIS}$ purely induced by the SOC. A negative $E_{BIS}$ is usually considered to be a feature of TI candidates, in Fig. 4(a)-(d) we display the region with the shadow yellow color. Let us first unveil the effects upon $E_{BIS}$ of hydrostatic strain. For the AgI, as shown in Fig. 4(a), the energy values of $E_{BIS}^{noSOC}$ and $E_{BIS}^{SOC}$ exhibit little fluctuation and stay above the zero energy level within the range of $a/a_0$ from 0.80 to 1.25, which indicates AgI an robust trivial phase upon hydrostatic strain. Moreover, the two curves are almost parallel and product the nearly horizontal SOC-induced $\Delta E_{BIS}$ with the value of ~0.25eV (most of the SOC effect originates from the core electrons, nearly unaffected during strain process). However, as shown in Fig. 4(c), the hydrostatic strain dependent $E_{BIS}$ exhibit distinct behaviours in AuI. Both the $E_{BIS}^{noSOC}$ and $E_{BIS}^{SOC}$ curves keep gradually climbing as the $a/a_0$ ratio rise, and present a transition from TI to a trivial phase at $a/a_0$=1.2. Such a transition upon strain, where $E_{BIS}$ occurs a sign change after switching on the SOC will be further discussed latter. Meanwhile the two climbing curves are also approximately equidistant, and give a value of $\Delta E_{BIS}$ ~0.20eV for AuI, which is even lower than that of AgI. This indicate the AuI system possess a weaker SOC strength, even though the Au is much heavier than Ag. Such an abnormal behavior can be ascribed to the lack of inversion symmetry in wurtzite structure. Simillar situation has been already discussed in CdTe and HgTe system in ref. 18 and ref. 32. The value of $E_{BIS}$ can be attributed to two factors: the SR effect [procedure (I) + procedure (II) in Fig. 3] and SOC [procedure (III) in Fig. 3]. However, the wurtzite AuI with a weaker SOC strength exhibits a naturally inverted band while AgI shows a normal one, the reason is that the heavier AuI system possess a much stronger SR effect than AgI. According to ref. 18 and ref. 33, the mass-velocity is extremly prominent for S electrons in heavy atoms, and cause a net downward band shift after the compensation of upward Darwin shift. This also illuminates the facts that most current TIs are found in heavy atoms containing compounds.

We next consider the impact of uniaxial strain on the $E_{BIS}$ for the wurtzite AgI and AuI. Encouragingly, the robust trivial AgI bubbles up under uniaxial strain as shown in Fig. 4(b), the $E_{BIS}^{noSOC}$ and $E_{BIS}^{SOC}$ gradually decrease with reducing $a/a_0$, and change their sign at $a/a_0$=0.86. It indicates the AgI will transform from the trivial phase to topological phase at a/a0=0.86. As to AuI under uniaxial strain in Fig. 4(d), The $E_{BIS}^{noSOC}$ and $E_{BIS}^{SOC}$ curves are almost upright with varing $a/a_0$ and get its transition point at $a/a_0$=1.02. We can gain more information by comparing the effects of hydrostatic and uniaxial strain for AgI [Fig. 4(a) and (b)] and AuI [Fig. 4(c) and (d)]. Compared with hydrostatic strain, the uniaxial one produces more dramatic impacts to the $E_{BIS}^{noSOC}$ and $E_{BIS}^{SOC}$, since the uniaxial strain leads to a more drastic lattice distortion and results strong change of hybridization strength. On this consideration, we can fruit TIs from the purely light elements containing systems via proper uniaxial strain, just as the layerd GaS and GaSe.[19] Moreover, the uniaxial strain also palys as an effective approach to open up a globle band gap in some cubic gapless TIs.[13,34]

Above discussions have confirmed that, the SOC is not the root cause but can add further fuel to the band inversion. It thus exists such situation that topological phase transition is opened by the SOC impetus, just as the cases of the critical points in Fig. 4(b)-(d). We just pick out the circled states in Fig. 4(c) for futher discussion, the corresponding band evolution diagram and electronic band structures are shown in Fig. 5. A 20% stretch to the lattice constant will weaken the SR effects in AuI system, then the downward shift of the $S_{Au^+}$ orbital will be not strong enough to trigger a band inversion after successively taking into account the effects of chemical bonding [procedure (I)] and crystal field splitting [procedure (II)]. The corresponding band structure from calculation without SOC shows an ordinary band order ($\Gamma_6 \rightarrow \Gamma_8 \rightarrow \Gamma_7$ from top to bottom). In procedure III, the $S_{Au^+}$ energy level steps over the threshold and sits below the $P_{I,xy,-}$ one after taking into account the SOC, which can be confirmed by the resulting band structure with SOC. Since the SOC-induced $E_{BIS}$ is limited, the s-like bands with $\Gamma_6$ symmetry just sit below $\Gamma_8$ but above the $\Gamma_7$ ones. This is distinct from the unstrained AuI case, where the $\Gamma_6$ energy level is lower. However, such difference would not affect the topological character, for the $P_{I,xy,-}$ and $P_{I,z,-}$ hold the same parity. Thus we obtain a topological nontrival phase with the SOC included.

## IV. CONCLUSION

Using the first-principle calculations, we predict that some of the hexagonal wurtzite compounds can serve as topological insulator candidates. Both the hypothetical AuI and strained-AgI exhibit inverted band behaviour between $S_+$ and $P_{xy,-}$ states, which are similiar with the cubic HgTe case. With the help of the caculated band structures and corresponding evolution schematics, we confirm that the scalar relativistic effects are essential for the band inversion mechanism rather than the the spin-orbit coupling. However, the latter is responsible to the globle band gap between the $\Gamma_{8+}$ and $\Gamma_{8-}$ states. We also find that the topological order can be turned by hydrostatic and uniaxial strains, but the uniaxial one performs a higher efficiency. Our work can be meaningful to widen the circle of topological insulator candidates.

This work is supported by the National Basic Research Program of China (973 Program 2012CB619405) and National Natural Science Foundation of China (Grant Nos. 51071172, 51171207 and 51021061)

**Figure captions:**

Fig. 1. Comparison of the zinc-blende and wurtzite crystal structures. The three-dimensional zinc blende and wurtzite structures are shown in (a) and (b), respectively. It is also shown that the cubic zinc-blende supercell could exhibit the sight of hexagonal lattice along the (111) direction [see Fig. 1(c)], which is comparable to the top side view of wurtzite structure in (d).

Fig. 2. Band structures of (a) CdTe and (d) HgTe compared with those of wurtzite compounds (b)AgI and (e)AuI with SOC included. The $\Gamma_6$ , $\Gamma_7$ and $\Gamma_8$ states are denoted by red, green and blue lines, respectively. To expose the effects of SOC, the band structures of AgI and AuI without SOC are also shown in (c) and (f).

Fig. 3. (a) Three procedures of the band evolution originating from the atomic S and $P_{x,y,z}$ orbitals at the $\Gamma$ point for CdTe. The pink dashed line marks the Fermi level. Procedure I represent the effects of introducing chemical bonding. Then in procedure II the cubic crystal field is taken into account, which push down the s and $p_{x,y,z}$ states. Finally the SOC is switched on in procedure III, where corresponding energy levels splited. (b), (c) and (d) are analogous to (a) but for AgI, HgTe and AuI, respectively.

Fig. 4. (a) The hydrostatic and (b) uniaxial strain dependent $E_{BIS}$ as a function of the $a/a_0$ ratio for AgI. $E_{BIS}^{noSOC}$ and $E_{BIS}^{SOC}$ are respectively represent the calculated $E_{BIS}$ without and with SOC. The energy discrepancy of $\Delta E_{BIS} = E_{BIS}^{noSOC} - E_{BIS}^{SOC}$ is also shown. A negative $E_{BIS}$ is usually considered to be a feature of TI candidates, and we display the region with the shadow yellow color. Similarly, the schemes of AuI for hydrostatic and uniaxial strain are shown in (c) and (d), respectively.

Fig. 5. Schematic diagram of the bande volution at the $\Gamma$ point for AuI under a 1.2% hydrostatic strain. The products of corresponding band structures without and with SOC are also shown below the schematic diagram.

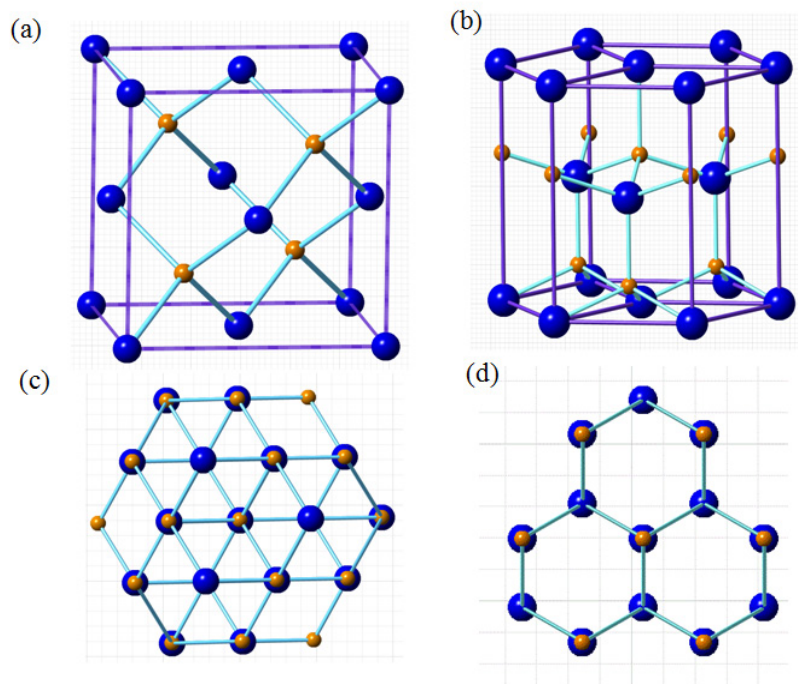

FIG. 1. Zhang et al.,

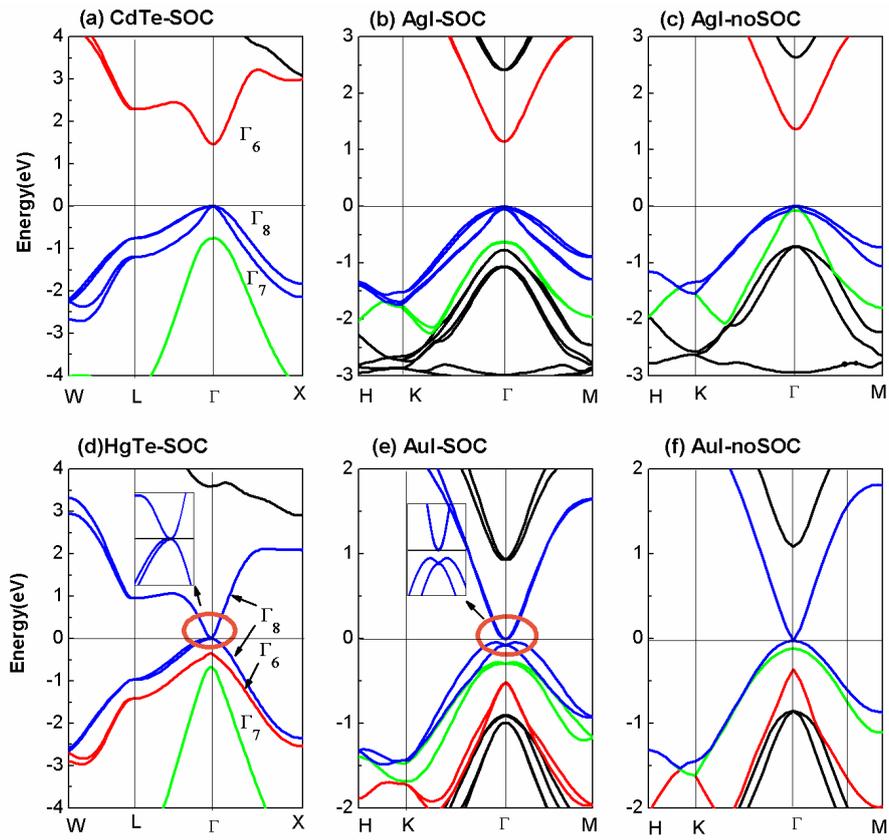

FIG. 2. Zhang et al.,

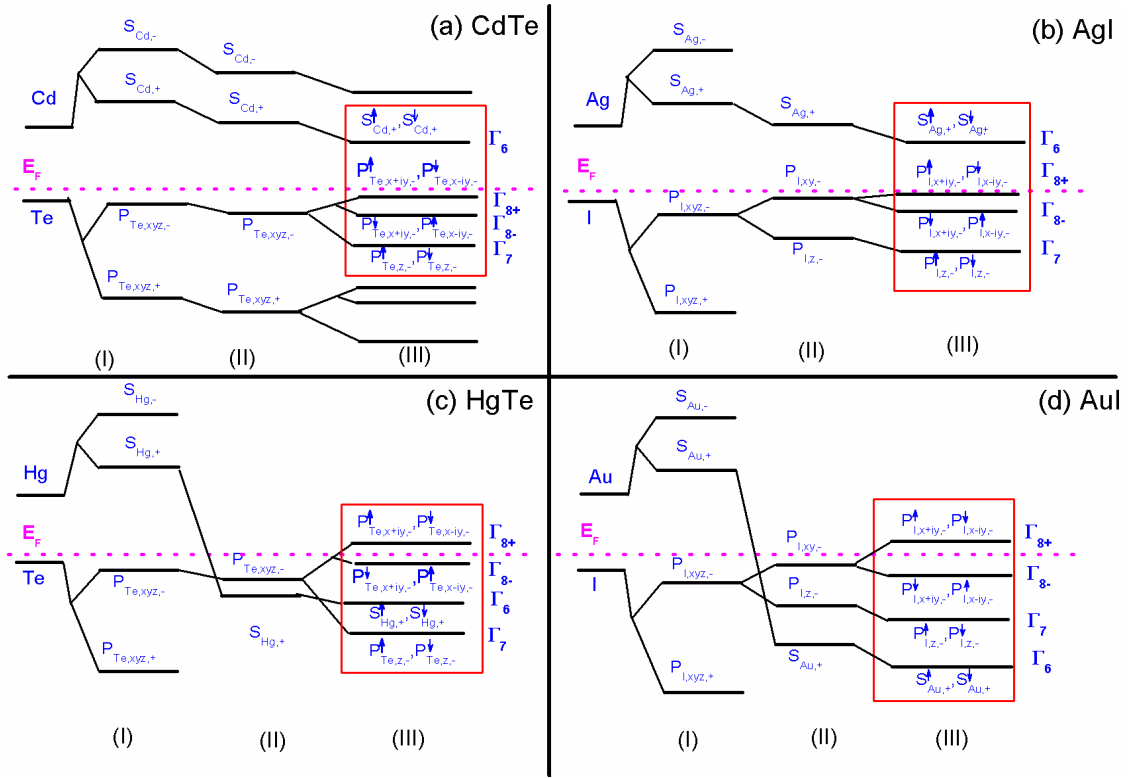

FIG. 3. Zhang et al.,

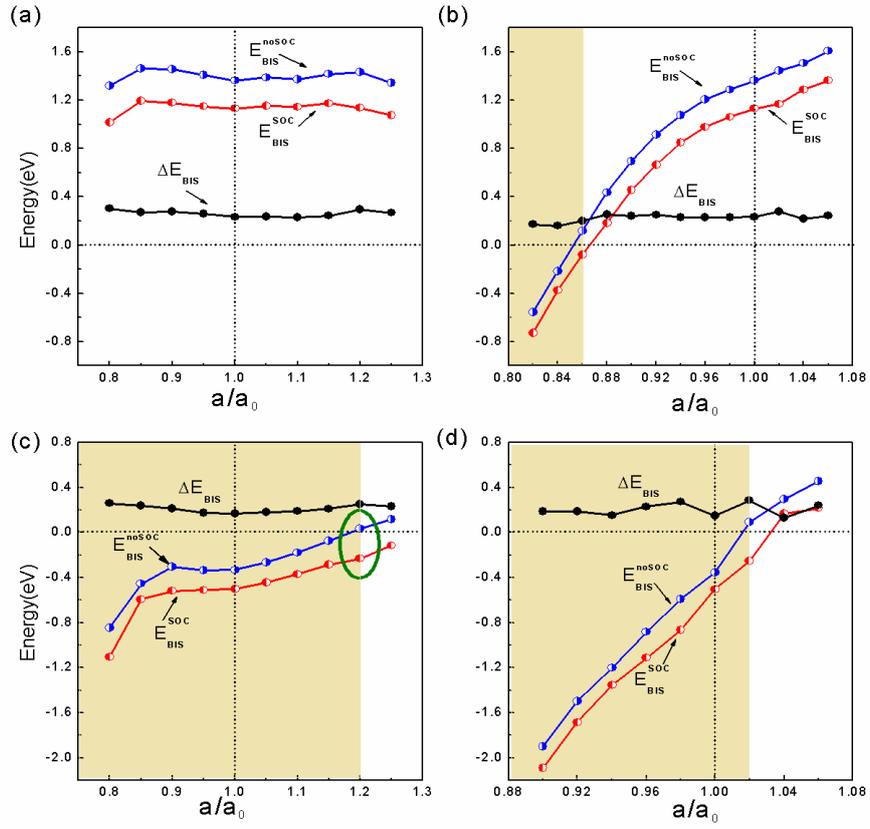

FIG. 4. Zhang et al.,

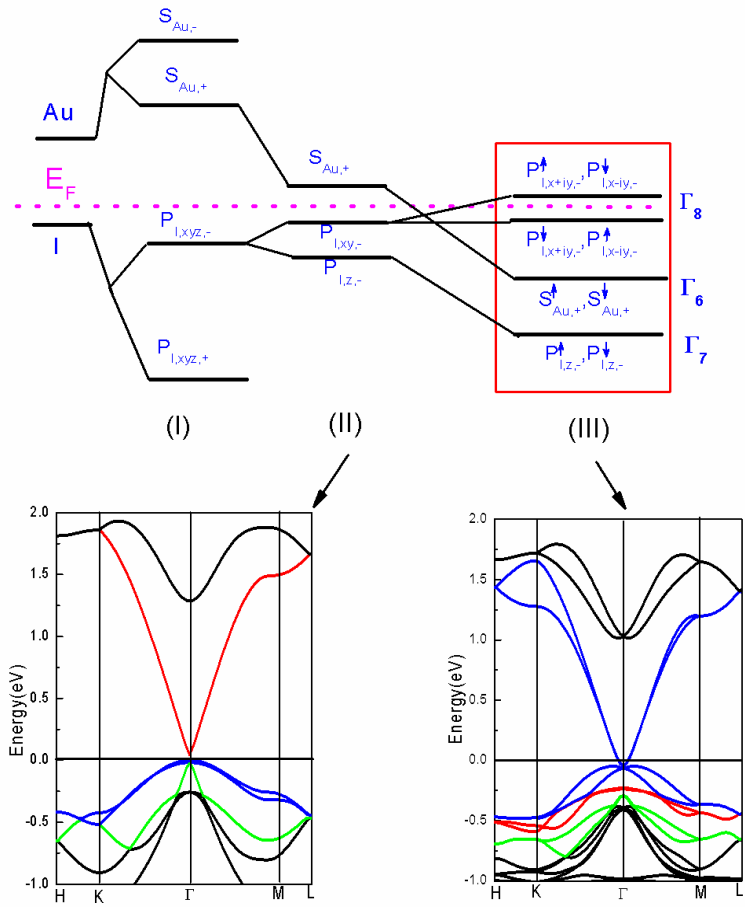

FIG. 5. Zhang et al.,